\documentclass[aps,prl,twocolumn,nopacs,superscriptaddress]{revtex4-2}

\usepackage[dvipsnames]{xcolor}
\usepackage{graphicx}
\usepackage{dcolumn}
\usepackage{bm}
\usepackage{hyperref}

\usepackage{braket}
\usepackage[utf8]{inputenc} 
\usepackage{mathtools}
\usepackage[bottom]{footmisc}
\usepackage{hyperref}
\usepackage{amsfonts}
\usepackage{amsmath}
\usepackage{slashed}
\usepackage{amsthm}

\begin{document}


\title{Generation of isolated flat bands with tunable numbers through Moiré engineering}

\author{Xiaoting Zhou}
\thanks{X. Z., Y. H., and B. W. contributed equally to this work.}
\email{x.zhou@northeastern.edu}
\affiliation{ Department of Physics,\;Northeastern\;University,\;Boston,\;Massachusetts,\;02115,\;USA}

\author{Yi-Chun Hung}
\thanks{X. Z., Y. H., and B. W. contributed equally to this work.}
\affiliation{ Department of Physics,\;Northeastern\;University,\;Boston,\;Massachusetts,\;02115,\;USA}

\author{ Baokai Wang}
\thanks{X. Z.,  Y. H., and B. W. contributed equally to this work.}
\affiliation{ Department of Physics,\;Northeastern\;University,\;Boston,\;Massachusetts,\;02115,\;USA}

\author{Arun Bansil}
\email{ar.bansil@northeastern.edu}
\affiliation{ Department of Physics,\;Northeastern\;University,\;Boston,\;Massachusetts,\;02115,\;USA}


\begin{abstract}
Unlike the spin-1/2 fermions, the Lieb and Dice lattices both host triply-degenerate low-energy excitations. Here we discuss Moiré structures involving twisted bilayers of these lattices, which are shown to exhibit a tunable number of isolated flat bands near the Fermi level. These flat bands remain isolated from the high-energy bands even in the presence of small higher-order terms and chiral-symmetry-breaking interlayer tunneling. At small twist angles, thousands of flat bands can be generated to substantially amplify flat band physics. We demonstrate that these flat bands carry substantial quantum weight so that upon adding a BCS-type pairing potential, the associated superfluid weight would also be large, and the critical superconducting temperature would be tunable. Our study suggests a new pathway for flat-band engineering based on twisted bilayer Lieb and Dice lattices. 
\end{abstract}

\maketitle
\paragraph*{\textbf{Introduction.---}}
\par Dispersionless electrons, commonly referred to as flat bands, provide a foundation for exploring strongly correlated physics. Flat bands generate high densities of states (DOS). As a result, the kinetic energy of the associated carriers is suppressed, and the interaction energy begins to dominate to trigger various correlated physical phenomena, especially when the flat bands are isolated near the Fermi energy. \cite{PhysRevB.34.5208, Calugaru2022, wang2023flat}. Exploration of flat-band systems is thus of fundamental importance, and subject of much current interest. 

\par There are many mechanisms for generating flat bands. They can be induced by the geometry or symmetry of the lattice to produce the so-called compact localized states (CLS) \cite{A.Mielke_1991a, A.Mielke_1991b, A.Mielke_1993}. In certain bipartite lattices \cite{PhysRevB.34.5208, Calugaru2022, wang2023flat}, the wave functions are localized on certain sublattices due to destructive interference. Examples include Kagome \cite{10.1143/ptp/6.3.306}, Lieb \cite{ PhysRevLett.62.1201}, and Dice lattices \cite{ PhysRevB.34.5208}. Such geometry-induced flat bands can be isolated from the high-energy bands via effects of spin-orbit coupling, dimerization, anisotropic strain \cite{PhysRevB.106.155417}, or stacking \cite{PhysRevB.99.125131, PhysRevB.101.045131, PhysRevB.107.035421, PhysRevB.99.155124}. The Dice lattice also hosts pseudospin-$1$ low-energy states at high-symmetry points \cite{PhysRevB.108.075166, PhysRevB.108.075167}, and these have been realized experimentally realization in SrTiO$_3$/SrIrO$_3$/SrTiO$_3$ superlattice \cite{doi:10.7566/JPSJ.87.041006}, and LaAlO$_3$/SrTiO$_3$ (111) quantum wells \cite{PhysRevLett.111.126804, PhysRevB.102.045105}.

\par Another widely recognized mechanism involves the Moiré bands observed in twisted van der Waals systems, such as bilayer graphene \cite{PhysRevLett.122.106405, PhysRevLett.123.036401, parhizkar2023generic, Devakul2021}, which supports superconductivity and provides a tunable system for realizing correlated phases \cite{ Cao2018, doi:10.1126/science.aav1910}. However, all known Moiré materials possess a fixed number of flat bands near the Fermi level, and their physics is controlled by the bandwidth that varies with twist angle \cite{PhysRevLett.122.086402, PhysRevLett.122.106405, PhysRevLett.123.036401, parhizkar2023generic, Devakul2021, doi:10.1073/pnas.1108174108}.

\par Here, we combine the two aforementioned mechanisms for generating flat bands and discuss properties of Moiré structures based twisted bilayer bipartite lattices, especially the Lieb and Dice lattices.
These Moiré structures are found to manifest a tunable number of isolated flat bands depending on the twist angle. We elucidate the origin of flat bands through an analysis of continuum models, where the valley structure is generated by the inclusion of second nearest neighbor (SNN) hoppings. Our conclusions are further substantiated by tight-binding calculations involving the absence of SNN hoppings in the twisted bilayer lattice. \\


\begin{figure}[h]
\begin{centering}
\includegraphics[width=\linewidth]{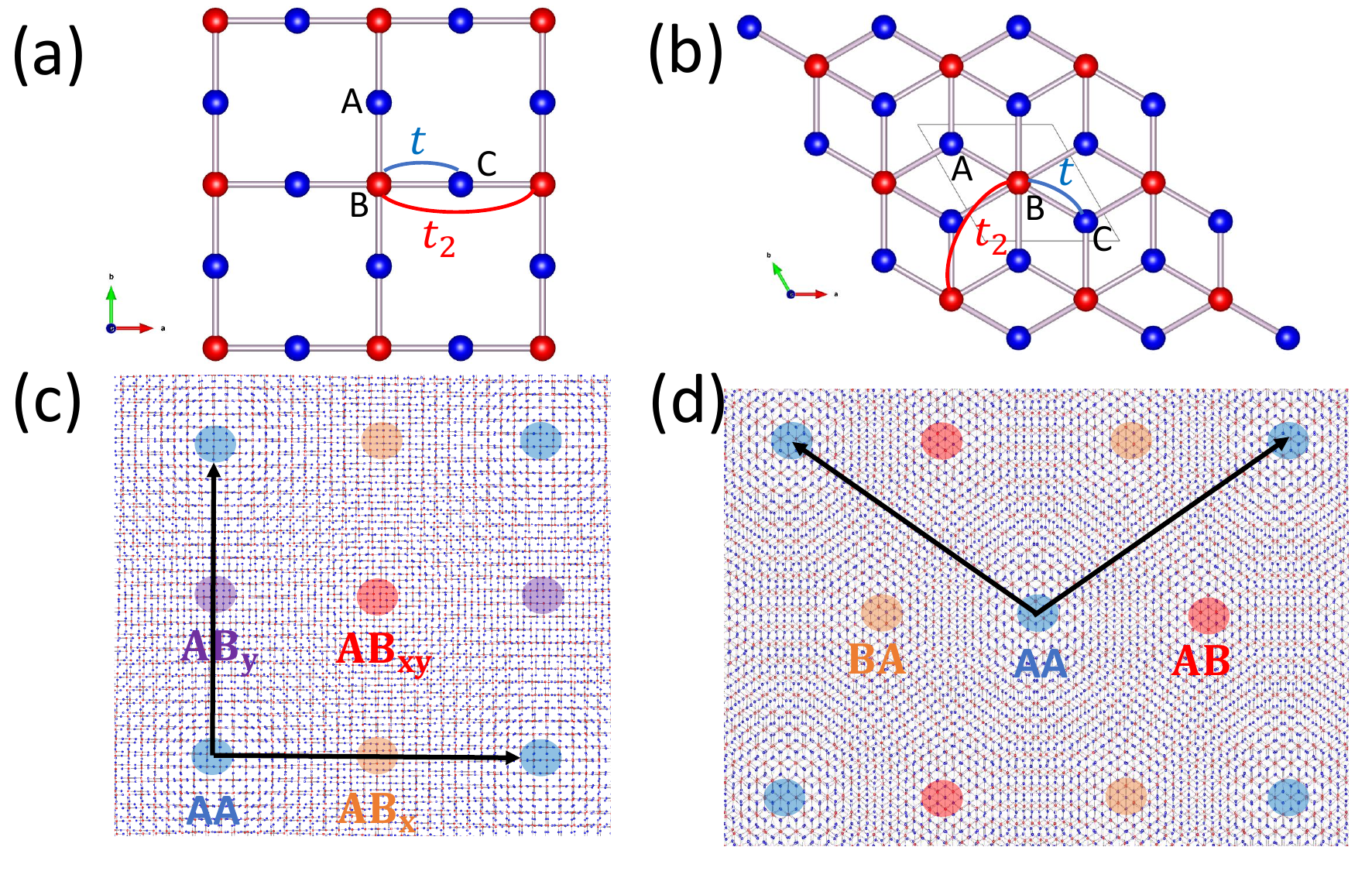}
\end{centering}
\centering{}\caption{ A schematic diagram of the Lieb (\textbf{a}) and Dice (\textbf{b}) lattices, in which the nearest-neighbor hopping $t$ is marked in blue and the second nearest-neighbor hopping $t_2$ is marked in red. The Moiré pattern of the twisted bilayer Lieb (\textbf{c}) and Dice (\textbf{d}) lattices. Centers of the high-symmetry stacked regions are marked by colored circles. Black arrows give the translation vectors of the Moiré unit cell. }
\label{fig:01}
\end{figure}

\paragraph*{\textbf{Twisted bilayer Lieb/Dice lattice.---}}
\par
\begin{figure}[h]
\begin{centering}
\includegraphics[width=\linewidth]{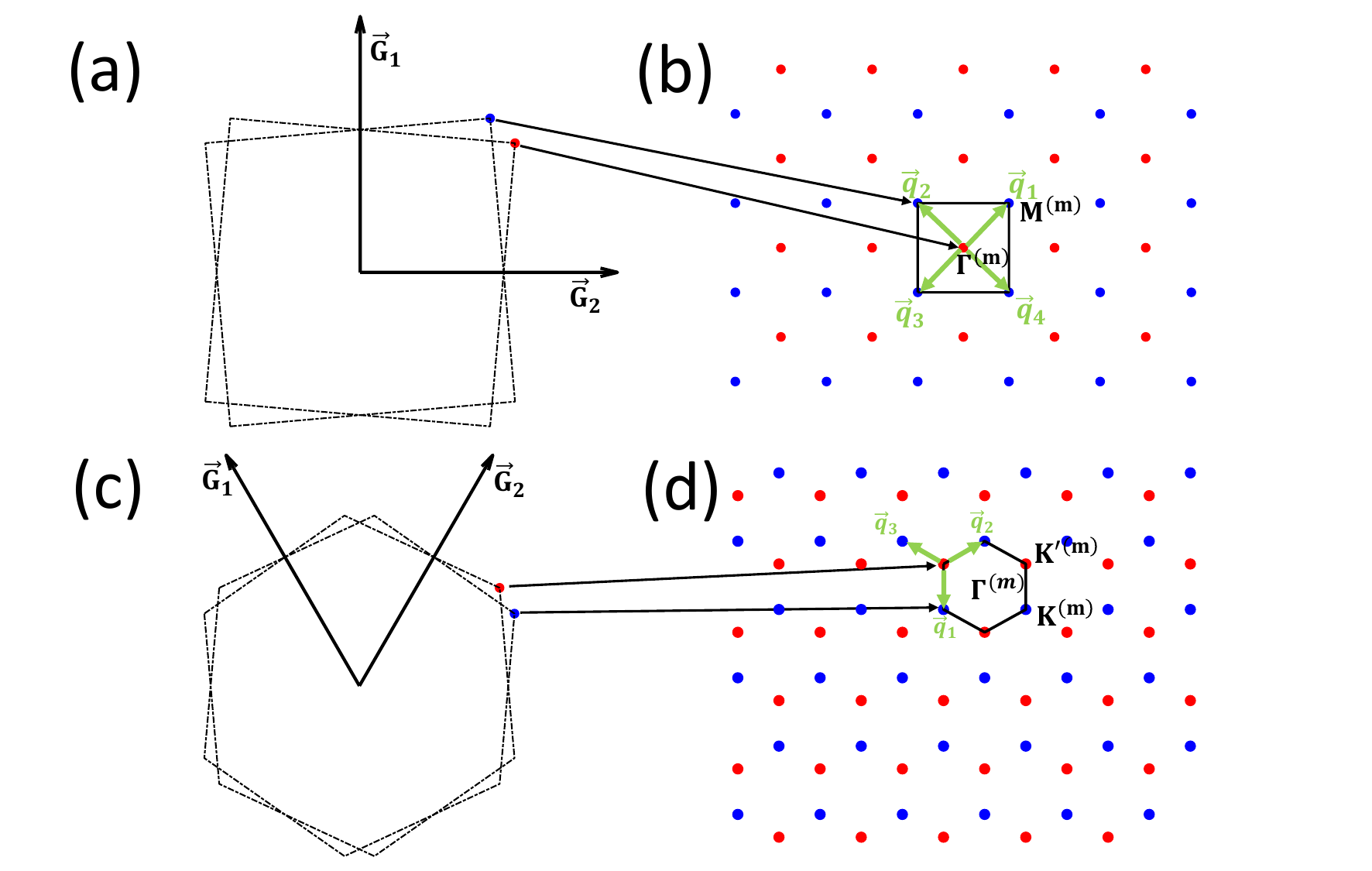}
\end{centering}
\centering{}\caption{(\textbf{a}) BZ of the Lieb lattice on the top (blue) and bottom (red) layer. $\vec{G}_i$ are the reciprocal lattice vectors for an unrotated monolayer Lieb lattice. (\textbf{b}) Moiré BZ of the TBL. The $M$-points of the top layer (blue) and of the bottom layer (red) map to the $M^{(m)}$ point and the $\Gamma^{(m)}$ point in the Moiré Brillouin zone, respectively. (\textbf{c}) The Brillouin zone of the Dice lattice on the top (blue) and bottom (red) layer. The $\vec{G}_i$ indicates the reciprocal lattice vectors of an unrotated monolayer Dice lattice. (\textbf{d}) The Moiré Brillouin zone of the TBD system. The $K$-points of the top layer (blue) and of the bottom layer (red) map to the $K'^{(m)}$ point and the $K^{(m)}$ point in the Moiré Brillouin zone, respectively. }
\label{fig:02}
\end{figure}

 We consider the generalized Lieb or Dice lattice as an unrotated monolayer system in which the second nearest neighbor hoppings are included (Figs.~\ref{fig:01}(a) and \ref{fig:01}(b)). With the second nearest-neighbor hopping $t_2$, the valley structures develop at the $M$ point in the Lieb and at $K$ and $K'$ in the Dice lattice, allowing low-energy expansions around these triply degenerate points, see Supplementary Material. Continuum models for the twisted bilayer Lieb (TBL) and twisted bilayer Dice (TBD) lattices can be obtained straightforwardly by following the Bistritzer-MacDonald formalism \cite{doi:10.1073/pnas.1108174108}:
\begin{equation}\label{eq:03}
    H(\theta) = \begin{pmatrix} H_0(\frac{\theta}{2}) & T \\ T^{\dagger} & H_0(-\frac{\theta}{2}) \end{pmatrix}.
\end{equation}
Here, for the TBL:
\begin{equation}\label{eq:03.5}
    H_0(\theta) = H_0^{(\text{Lieb})}(\theta) = t((\tilde{q}_x)\lambda_2 + (\tilde{q}_y)\lambda_7) + t_2q^2,
\end{equation}
where $\tilde{q}_x\equiv q_x\cos(\theta)- q_y\sin(\theta)$, $\tilde{q}_y\equiv q_x\sin(\theta)+q_y\cos(\theta)$, and $\vec{q}=\vec{k}-\vec{k}_{M}^{(\theta)}$ with $\vec{k}_{M}^{(\theta)}$ be the $k$-vector of the $M$ valley in the \emph{rotated} monolayer Brillouin zone (FIG.~\ref{fig:02}(a)). The $\{\lambda_i\}$ are Gellmann matrices \cite{Gellmann}. On the other hand, for the TBD:
\begin{equation}\label{eq:13}
    H_0(\theta) = H_0^{(\text{Dice})}(\theta,\zeta) = \frac{3t}{2}(\zeta q_x\hat{x}-q_y\hat{y})\cdot \vec{S}^{(\theta)} + t_2q^2,
\end{equation}
where $\vec{S}^{(\theta)}\equiv e^{-i\frac{\theta}{2}S_3}\vec{S}e^{i\frac{\theta}{2}S_3}$ and $\zeta=\pm$ indicates the valley degree of freedom. The vector $\vec{q}=\vec{k}-K_{\zeta}^{(\theta)}$ with $K_{\zeta}^{(\theta)}$ be the $K_\zeta$ valley in the \emph{rotated} monolayer BZ of Dice lattice ( FIG.~\ref{fig:02}(b)). The $\{S_i\}$ are matrix representations of spin-$1$ in the basis of $s_z$ eigenstates \cite{spin1}. Note that the basis in Eqs.~\ref{eq:03.5} and \ref{eq:13} is $\begin{pmatrix} A, & B, & C\end{pmatrix}^T$, where $A$, $B$, and $C$ are sublattices in the monolayer lattices (FIG.~\ref{fig:01}).

The interlayer coupling $T$ has the following form \cite{doi:10.1073/pnas.1108174108, PhysRevB.99.155415}:
\begin{equation}\label{eq:04}
    T = \begin{pmatrix} \omega_1g(\vec{r}) & \omega_2g(\vec{r}-\vec{r}_{12}) & \omega_3g(\vec{r}-\vec{r}_{13}) \\ 
         \omega_2g(\vec{r}+\vec{r}_{12}) & \omega_1g(\vec{r}) & \omega_2g(\vec{r}-\vec{r}_{23}) \\ 
         \omega_3g(\vec{r}+\vec{r}_{13}) & \omega_2g(\vec{r}+\vec{r}_{23}) & \omega_1g(\vec{r}) \end{pmatrix},
\end{equation}
where $g(\vec{r})=\sum_ie^{i\vec{q}_i\cdot\vec{r}}$ with $\vec{q}_i$ representing the momentum transfer of interlayer tunneling between valleys from different layers in the Moiré Brillouin zone ( Figs.~\ref{fig:02}(a) and \ref{fig:02}(b)). For TBL, $\vec{q}_1 = \frac{\pi}{L_s}\hat{x} + \frac{\pi}{L_s}\hat{y}$, $\vec{q}_2 = -\frac{\pi}{L_s}\hat{x} + \frac{\pi}{L_s}\hat{y}$, $\vec{q}_3 = \frac{\pi}{L_s}\hat{x} - \frac{\pi}{L_s}\hat{y}$, and $\vec{q}_4 = -\frac{\pi}{L_s}\hat{x} - \frac{\pi}{L_s}\hat{y}$. For TBD, $\vec{q}_1 = -\frac{4\pi}{3L_s}\hat{y}$, $\vec{q}_2 = \frac{2\pi}{\sqrt{3}L_s}\hat{x} + \frac{2\pi}{3L_s}\hat{y}$, $\vec{q}_3 = -\frac{2\pi}{\sqrt{3}L_s}\hat{x} + \frac{2\pi}{3L_s}\hat{y}$. In Eq.~\ref{eq:04}, $\vec{r}_{ij}$ represents the center of different stacked regions in the Moiré unit cell, and the origin of the coordinates is set at the center of the AA-stacked region, see Supplementary Material for details of lattice geometry of high-symmetry stacking. For the TBL, positions of different stacked regions are $\vec{r}_{12}=\vec{r}_{AB_x}=\frac{L_s}{2}\hat{y}$, $\vec{r}_{23}=\vec{r}_{AB_y}=\frac{L_s}{2}\hat{x}$, and $\vec{r}_{13}=\vec{r}_{AB_{xy}}=\frac{L_s}{2}(\hat{x}+\hat{y})$ (see FIG.~\ref{fig:01}(a)). For the TBD, positions of different stacked regions are $\vec{r}_{12}=\vec{r}_{23}=-\vec{r}_{13}=\vec{r}_{AB}=\frac{L_s}{\sqrt{3}}\hat{x}$ ( FIG.~\ref{fig:01}(b)). \\

\paragraph*{\textbf{Electronic structure of twisted bilayers.---}}
\par
\begin{figure}[ht]
\begin{centering}
\includegraphics[width=\linewidth]{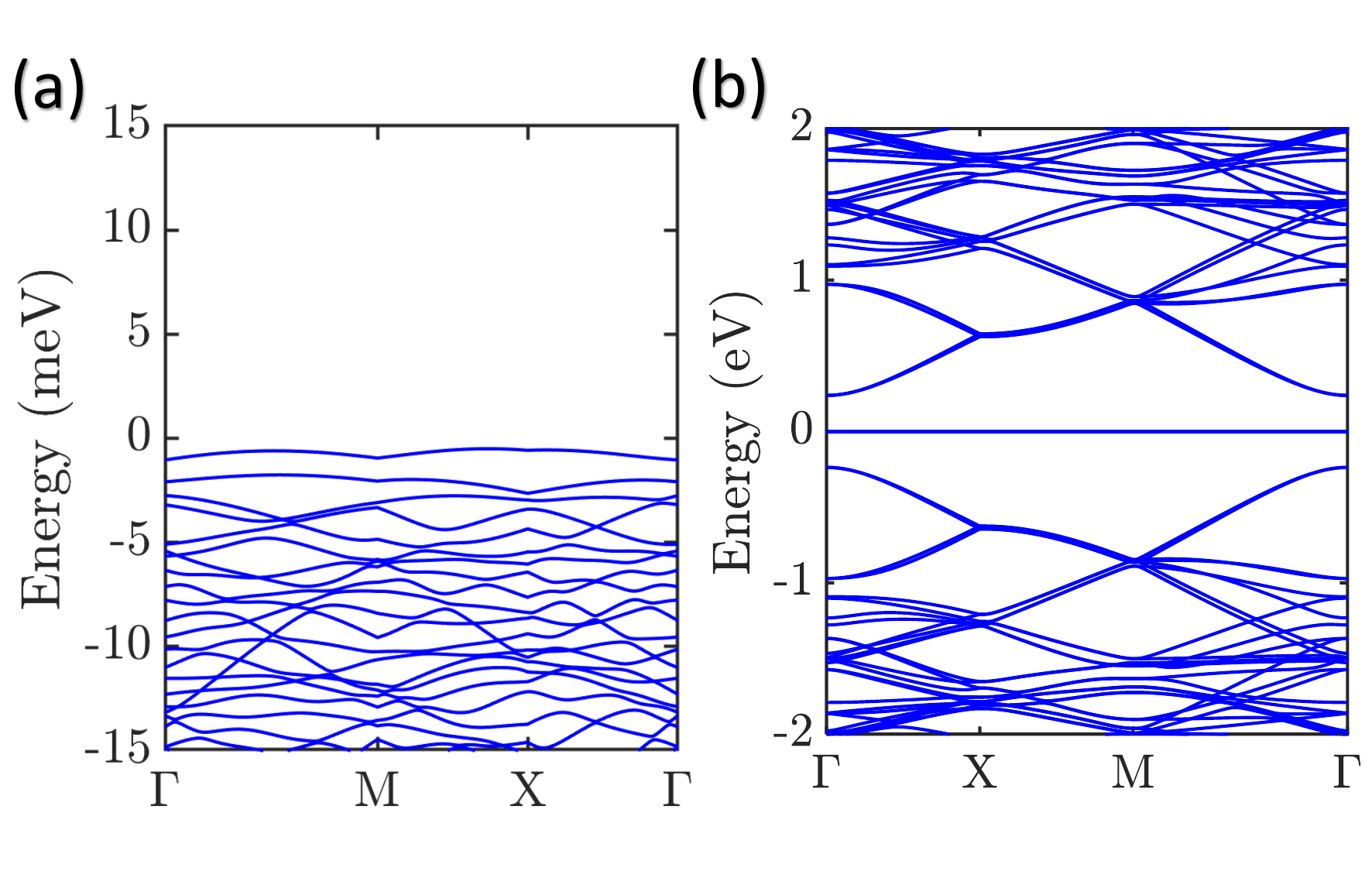}
\end{centering}
\centering{}\caption{(\textbf{a}) The band structure of the Hamiltonian in Eq.~\ref{eq:03} for TBL with $\alpha\cong1.423$, in which $t=5.817$ eV, $t_2=-0.1$ eV, $\omega_1=-0.001$ eV, $\omega_2\cong1.163$ eV, and $\omega_3=0$.
(\textbf{b}) The band structure of the TBL generated by the tight-binding model with Slater-Koster parameterization, which corresponds to the continuum model in Eq.~\ref{eq:03} with $\alpha\cong1.423$ and $t_2=\omega_1=\omega_3=0$. }
\label{fig:03}
\end{figure}

\begin{figure}[ht]
\begin{centering}
\includegraphics[width=\linewidth]{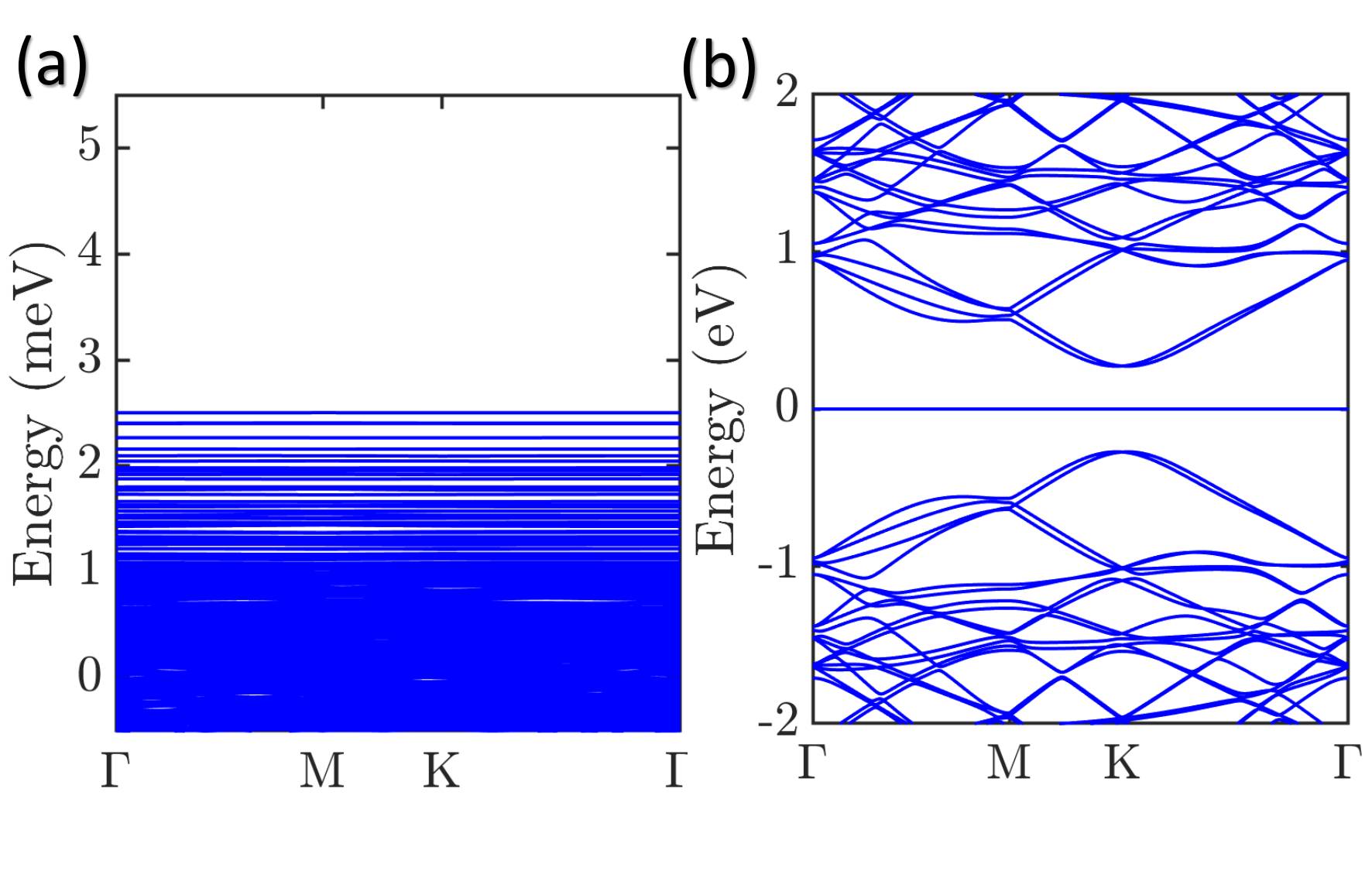}
\end{centering}
\centering{}\caption{(\textbf{a}) The band structure of the Hamiltonian in Eq.~\ref{eq:03} for TBD with $\alpha\cong19.429$, in which $t=5.817$ eV, $t_2=-0.1$ eV, $\omega_1=0.001$ eV, $\omega_2\cong1.163426$ eV, and $\omega_3=0$ eV.
(\textbf{b}) The band structure of the TBD generated by the tight-binding model with Slater-Koster parameterization, which corresponds to the continuum model in Eq.~\ref{eq:03} with $\alpha\cong19.429$ and $t_2=\omega_1=\omega_3=0$.}
\label{fig:04}
\end{figure}

 The band structures obtained from diagonalizing Eq.~\ref{eq:03} in the plane-wave basis are shown in FIG.~\ref{fig:03}(a.) for TBL and FIG.~\ref{fig:04}(a.) for TBD, which show isolated flat bands at the band edge around the Fermi level. These isolated flat bands are generated by gapping the folded parabolic dispersion originating from the second nearest-neighbor hopping through the interlayer tunneling $\omega_1$. To demonstrate this fact and get more insight into the TBL and TBD, let's consider the case where $\omega_1=\omega_3=t_2=0$. In this scenario, the electrons hop only between the $A,B$ sublattice and $A,C$ sublattice in both layers, which keeps the twisted bilayer lattice bipartite \cite{PhysRevB.34.5208, Calugaru2022, wang2023flat}, making the Hamiltonian in Eq.~\ref{eq:03} non-invertible and, therefore, hosting robust zero-energy states \cite{YUCE20191791}. When $t_2=0$, the existence of flat bands poses a challenge in validating Moiré band structure calculations using low-energy continuum models. 

\par
\begin{figure}[ht]
\begin{centering}
\includegraphics[width=\linewidth]{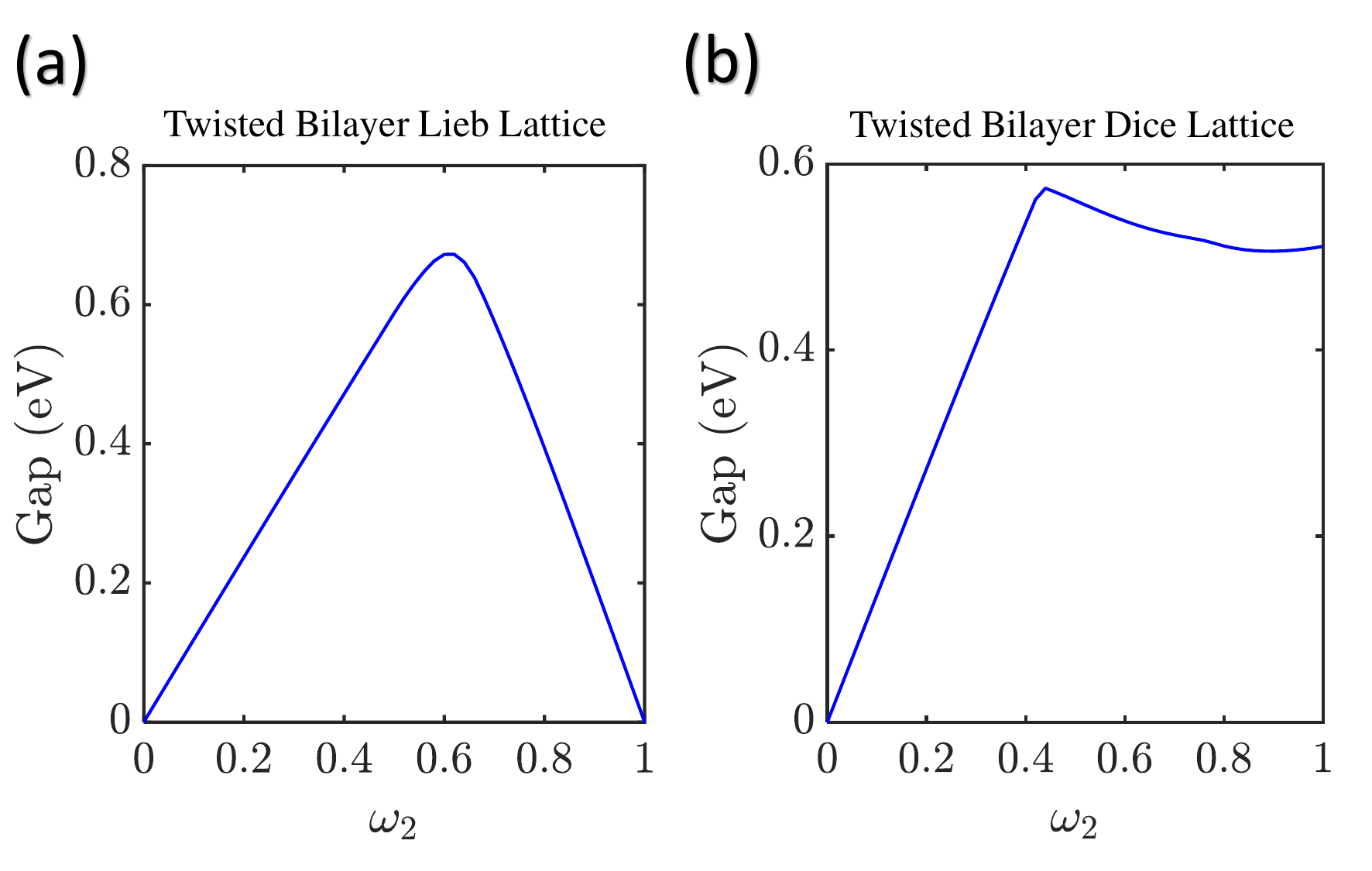}
\end{centering}
\centering{}\caption{(\textbf{a}) The gap between the flat bands and other high-energy bands in the tight-binding model of TBL with $t_2=\omega_1=\omega_3=0$ as a function of interlayer tunneling $\omega_2$.
(\textbf{b}) The gap between the flat bands and the high-energy bands in the tight-binding model of TBD with $t_2=\omega_1=\omega_3=0$ as a function of interlayer tunneling $\omega_2$. }
\label{fig:05}
\end{figure}

 To properly describe the TBL and TBD when $t_2=0$, we construct the tight-binding model with Slater-Koster parameterization. 
The setting $\omega_1=\omega_3=0$ corresponds to no interlayer tunneling between the same sublattice on different layers and no interlayer tunneling from $A$ and $C$ sublattices on different layers. When second-nearest neighbor hopping is not considered, the vanishing $\omega_1$ and $\omega_3$ leads to isolated exact flat bands at $E=0$ with considerable degeneracies in the Moiré Brillouin zone (see FIG.~\ref{fig:03}(b.) and FIG.~\ref{fig:04}(b.)). In this scenario, the system is controlled by the parameter $\alpha\equiv\frac{\omega_2}{tk_\theta}$. Due to the special hopping structure in TBD and TBL when $\omega_1=\omega_3=t_2=0$, the number of these flat bands is $\frac{1}{3}$ of the total bands. Remarkably, these flat bands remain isolated for a large range of $\alpha$ as shown in FIG.~\ref{fig:05}(a.) and FIG.~\ref{fig:05}(b.).

\par
\begin{figure}[h]
\begin{centering}
\includegraphics[width=\linewidth]{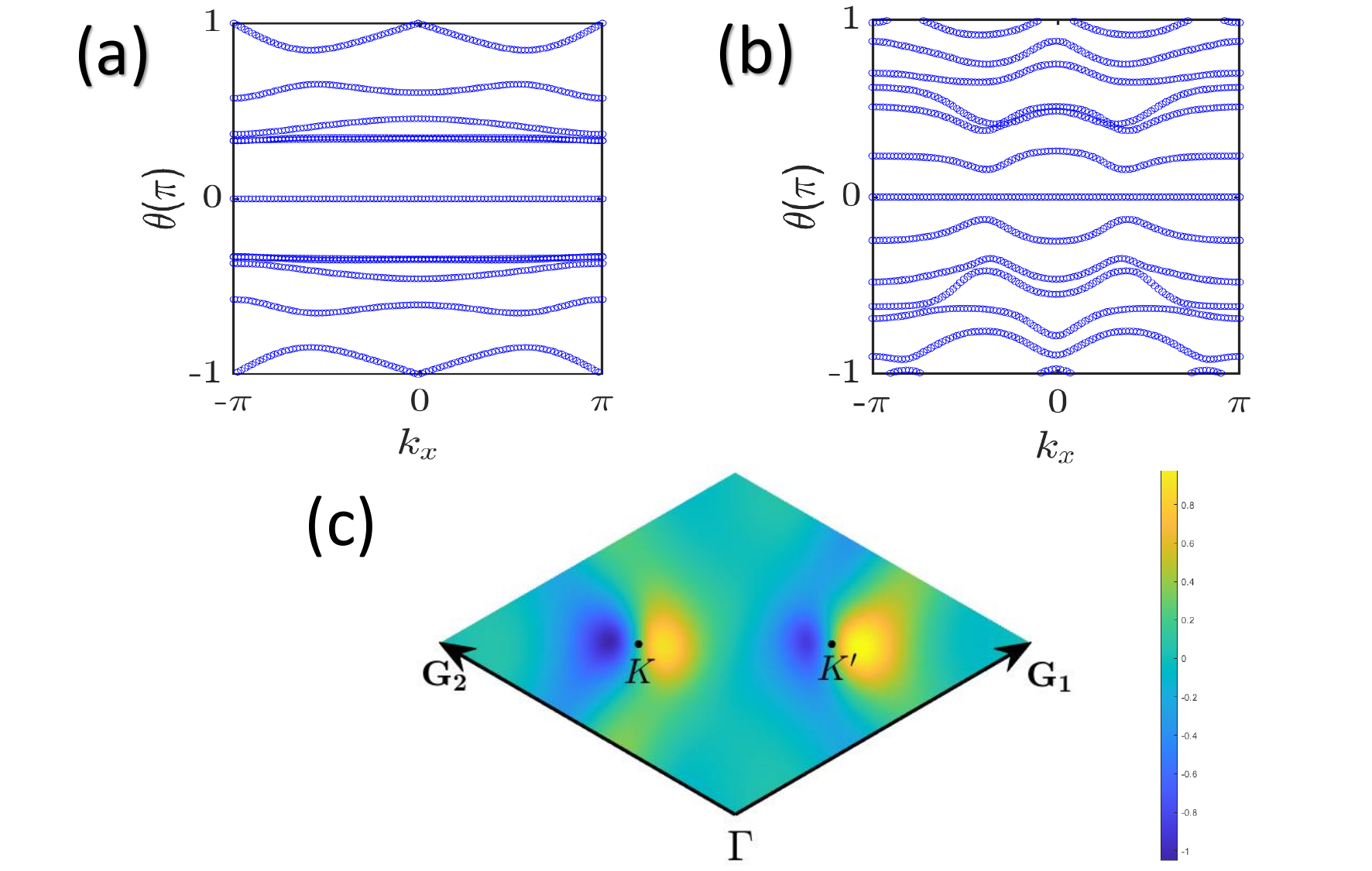}
\end{centering}
\centering{}\caption{(\textbf{a}) The hybridized Wannier center of the isolated flat bands in TBL. (\textbf{b}) The hybridized Wannier center of the isolated flat bands in TBD. (\textbf{c}) The Berry curvature generated by the isolated flat bands of the TBD at $\alpha\cong19.429$ in the Moiré Brillouin zone, which is concentrated around the valleys.}
\label{fig:06}
\end{figure}

 Due to the time-reversal symmetry, these isolated flat bands have no net Chern numbers. The Wilson loop calculation also shows that the isolated flat bands in TBL and TBD do not have non-trivial winding of the hybridized Wannier centers (see FIG.~\ref{fig:06}(a.) and FIG.~\ref{fig:06}(b.)). Due to the time-reversal symmetry, the flat bands in the TBL have no Berry curvature around its $M$ valley. In contrast, the flat bands in the TBD are found to host considerable Berry curvature around the $K,K'$ valleys (see FIG.~\ref{fig:06}(c.)). Note that the Berry curvature around each valley doesn't give a non-trivial valley Chern number, which is consistent with the analysis from the pseudo-Landau level representation of the isolated flat bands in TBD \cite{PhysRevB.99.155415}. More specifically, due to the bipartite nature, the pseudo-Landau level spectrum has $\frac{1}{3}$ total number of Landau levels are zero modes. However, we find that the zero-mode of the $n=0$ pseudo-Landau level cancels out the contribution of the Hall conductivity from the zero-modes of $n\neq0$ pseudo-Landau level (see Supplementary Material). In contrast to twisted bilayer graphene, the zero-mode generated by $n=0$ Landau levels in TBD do not contribute an integer conductivity quanta, and the zero-modes of the pseudo-Landau level in TBD are also generated by $n\neq0$ Landau levels, which makes the zero-modes of pseudo-Landau levels contribute no net Chern number. \\

\begin{figure}[h]
\begin{centering}
\includegraphics[width=\linewidth]{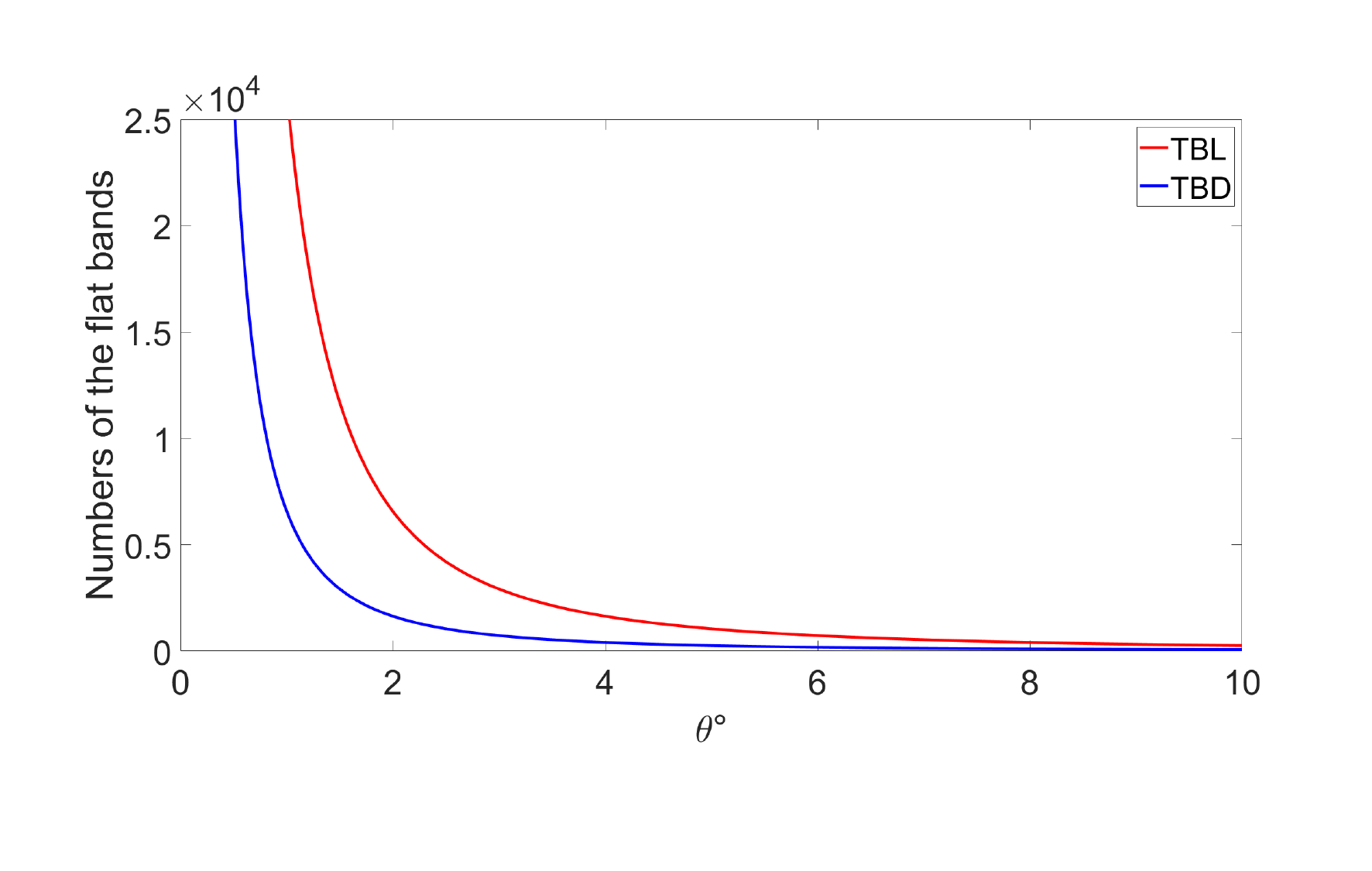}
\end{centering}
\centering{}\caption{The number of flat bands in TBL and TBD as a function of twisting angle $\theta$. When $\theta$ is small, the number of sublattice sites in the Moiré unit cell of TBL $N_{\text{sites}}^{\text{(TBL)}}\sim6\csc^2(\frac{\theta}{2})$ \cite{can_tummuru_day_elfimov_damascelli_franz_2021} and the number of flat bands in the Moiré unit cell of TBD $N_{\text{sites}}^{\text{(TBD)}}\sim\frac{3}{1-\cos(\theta)}$ \cite{PhysRevLett.99.256802}. The corresponding number of flat bands in TBL is $N_{\text{flat}}^{\text{(TBL)}}=\frac{1}{3}N_{\text{sites}}^{\text{(TBL)}}$ and in TBD is $N_{\text{flat}}^{\text{(TBD)}}=\frac{1}{3}N_{\text{sites}}^{\text{(TBD)}}$.}
\label{fig:07}
\end{figure}

\paragraph*{\textbf{Discussion---}}
\par In contrast to prior twisted materials where the count of low-energy flat bands remains constant with varying twisted angles, the twisted bilayer Lieb (TBL) and Dice (TBD) lattices exhibit a twist angle-dependent quantity of flat bands near the Fermi level. Notably, these quantities increase at a rate greater than $\frac{1}{\theta^2}$ as the twist angle $\theta$ decreases (FIG.~\ref{fig:07}).

\par The introduction of second nearest-neighbor (SNN) hopping $t_2$ and chiral-symmetry-breaking tunneling terms $\omega_1$ or $\omega_3$ disrupts the bipartite nature of the Moiré lattice structure, resulting in the coupling of these flat bands. However, as long as these effects are not overly significant, the cluster of flat bands near the Fermi level remains isolated from the high-energy spectra. Specifically, the consideration of SNN hopping $t_2$ leads to a parabolic dispersion of the flat bands near the valleys. Meanwhile, the inclusion of $\omega_1$ acts as an interaction, opening a gap in the parabolic dispersion and creating isolated flat bands at the upper band edge of the flat band cluster. In this scenario, the system is governed by three key parameters: $\alpha\equiv\frac{\omega_2}{tk_\theta}$, $\beta\equiv\frac{t_2}{tk_\theta}$, and $\gamma\equiv\frac{\omega_1}{tk_\theta}$. A similar effect is expected with the inclusion of $\omega_3$. Further exploration of the topological phase diagram concerning $\alpha$, $\beta$, and $\gamma$ will be interesting.

\par It is noteworthy that the twisted bilayer Dice lattice (TBD) exhibits a valley structure characterized by a concentrated Berry curvature, signifying a substantial quantum weight. Consequently, upon introducing a BCS-type pairing potential, this leads to a noteworthy superfluid weight \cite{PhysRevB.90.165139, PhysRevLett.124.167002, onishi2023fundamental}. Moreover, the dramatic variation in the density of states (DOS) with respect to the twist angle suggests a promising way for the realization of tuning the transition temperature by adjusting the twist angle. As the number of flat bands increases with a decrease in the twist angle, it is anticipated that the transition temperature will increase at a smaller twist angle due to the growing DOS \cite{grosso2013solid}. This paves the way toward controllable high-temperature superconductivity.

\par The count of flat bands near the Fermi level undergoes a dramatic change with varying twist angles, especially noticeable at small angles. Therefore, Moiré engineering of the TBL and TBD offers an exceptional platform for investigating strongly correlated physics, in which a slight tweak in the twist angle can exert a large influence on low-energy physics. For instance, manipulating the twist angle allows for engineering spontaneous ferromagnetic phase transitions based on the Stoner criterion. This is achieved by inducing dramatic changes in the DOS near the Fermi level, considering the spin degree of freedom through significant spin-orbit coupling or the presence of magnetic impurities \cite{grosso2013solid}. Notably, the TBL and TBD exemplify how Moiré engineering paves a novel pathway for manipulating geometry-induced flat bands in bipartite lattices.

\par The Moiré pattern provides not only an alternative way, distinct from dimerization or anisotropic strain, to isolate flat bands in bipartite lattices but also enables the creation of a large supercell, folding numerous flat bands into the first Brillouin zone. Additionally, while inheriting flat band physics from the unrotated monolayer, the Moiré pattern generates a distinctive electronic structure characterized by highly degenerate flat bands near the Fermi level. Further investigation into strongly correlated physics and the corresponding phase diagram within the realm of twisted bipartite lattices will be interesting.

\begin{acknowledgements}
\section*{Acknowledgements}
\par We thank Gregory Fiete for the discussions. The work was supported by the Air Force Office of Scientific Research under award number FA9550-20-1-0322 and benefited from the computational resources of Northeastern University’s Advanced Scientific Computation Center (ASCC) and the Discovery Cluster.
\end{acknowledgements}
\paragraph*{Note:} After the completion of our manuscript, we became aware of the related paper \cite{ma2023moire}.

 \bibliography{ref_aps}

\begin{thebibliography}{38}
\bibitem [{Gel()}]{Gellmann}%
  \BibitemOpen
  \href@noop {} {\bibinfo {title} {The gellmann matrices in this letter are:
  $\lambda_1=\begin{pmatrix}0 & 1 & 0\\ 1 & 0 & 0 \\ 0 & 0 & 0\end{pmatrix}$,
  $\lambda_2=\begin{pmatrix}0 & -i & 0\\ i & 0 & 0 \\ 0 & 0 &
  0\end{pmatrix}$,$\lambda_3=\begin{pmatrix} 1 & 0 & 0\\ 0 & -1 & 0 \\ 0 & 0 &
  0\end{pmatrix}$, $\lambda_4=\begin{pmatrix}0 & 0 & 1\\ 0 & 0 & 0 \\ 1 & 0 &
  0\end{pmatrix}$, $\lambda_5=\begin{pmatrix}0 & 0 & -i\\ 0 & 0 & 0 \\ i & 0 &
  0\end{pmatrix}$,$\lambda_6=\begin{pmatrix}0 & 0 & 0\\ 0 & 0 & 1 \\ 0 & 1 &
  0\end{pmatrix}$,$\lambda_7=\begin{pmatrix}0 & 0 & 0\\ 0 & 0 & -i \\ 0 & i &
  0\end{pmatrix}$,$\lambda_8=\frac{1}{\sqrt{3}}\begin{pmatrix}1 & 0 & 0\\ 0 & 1
  & 0 \\ 0 & 0 & -2\end{pmatrix}$}}\BibitemShut {NoStop}%
\bibitem [{spi()}]{spin1}%
  \BibitemOpen
  \href@noop {} {\bibinfo {title} {The spin matrices in this letter are:
  $s_1=\frac{1}{\sqrt{2}}\begin{pmatrix}0 & 1 & 0\\ 1 & 0 & 1 \\ 0 & 1 &
  0\end{pmatrix}$, $s_2=\frac{1}{\sqrt{2}}\begin{pmatrix}0 & -i & 0\\ i & 0 &
  -i \\ 0 & i & 0\end{pmatrix}$,$s_3=\begin{pmatrix} 1 & 0 & 0\\ 0 & 0 & 0 \\ 0
  & 0 & -1\end{pmatrix}$}}\BibitemShut {NoStop}%
\end{thebibliography}


\clearpage
\pagebreak
\widetext
\setcounter{equation}{0}
\setcounter{figure}{0}
\setcounter{table}{0}

\renewcommand{\theequation}{S\arabic{equation}}
\renewcommand{\thefigure}{S\arabic{figure}}
\renewcommand{\thetable}{S\arabic{table}}
\renewcommand{\bibnumfmt}[1]{[S#1]}
\renewcommand{\citenumfont}[1]{S#1}
\newcommand{\bk}{\boldsymbol\kappa}

\newcommand{\SI}{Supplementary Material}
\newcommand{\beginsupplement}{%
  \setcounter{equation}{0}
  \renewcommand{\theequation}{S\arabic{equation}}%
  \setcounter{table}{0}
  \renewcommand{\thetable}{S\arabic{table}}%
  \setcounter{figure}{0}
  \renewcommand{\thefigure}{S\arabic{figure}}%
  \setcounter{section}{0}
  \renewcommand{\thesection}{S\Roman{section}}%
  \setcounter{subsection}{0}
  \renewcommand{\thesubsection}{S\Roman{section}.\Alph{subsection}}%
}

\begin{center}
\textbf{\large Supplemental Materials of Generation of isolated flat bands with tunable numbers through Moiré engineering}
\end{center}
\section{The tight-binding model of the generalized Lieb lattice}
\par In the basis of $\begin{pmatrix} A, & B, & C\end{pmatrix}^T$, the Bloch Hamiltonian with only nearest-neighbor hopping on a Lieb lattice possessing sublattice symmetry is:
\begin{equation}\label{eq:01}
    H_{\text{bulk}}(\vec{k}) = t\begin{pmatrix} 0 & 1+e^{ik_x} & 0 \\ 1+e^{-ik_x} & 0 & 1+e^{ik_y} \\ 0 & 1+e^{-ik_y} & 0 \end{pmatrix},
\end{equation}
which has a triply degenerated touching at $M$ point and a flat band at $E=0$. Due to the flat band at $E=0$, describing the low-energy theory of the Lieb lattice through the continuum model is invalid. Fortunately, this problem can be solved by introducing the second nearest-neighbor hopping into the Lieb lattice. If the second nearest-neighbor hopping $t_2$ is considered, the Hamiltonian in Eq.~\ref{eq:01} becomes:
\begin{equation}\label{eq:09}
    H_{\text{bulk}}(\vec{k}) = t\begin{pmatrix} h & 1+e^{ik_x} & 0 \\ 1+e^{-ik_x} & h & 1+e^{ik_y} \\ 0 & 1+e^{-ik_y} & h \end{pmatrix},
\end{equation}
where $h=\frac{2t_2}{t}(\cos(k_x)+\cos(k_y))$. With the consideration of the second nearest-neighbor hopping, the flat band disperses, and the band structure develops a valley structure around $M$ point. Then, the low-energy Hamiltonian around $M$ can be described by:
\begin{equation}\label{eq:09.5}
    H_0(\vec{k}) = t(k_x\lambda_2 + k_y\lambda_7) + t_2k^2 - 4t_2,
\end{equation}
where $\lambda_i$ is the $i^{th}$ Gell-Mann matrix. Therefore, to rigorously analyze the low-energy theory of the moiré pattern in TBL, we study the low-energy theory in Eq.~\ref{eq:09.5} with $t_2\rightarrow0$ and perform calculation results through tight-binding Hamiltonian parameterized by the Slater-Koster method.

\begin{figure}[ht]
\centering
\includegraphics[width=8.6cm, height=5cm]{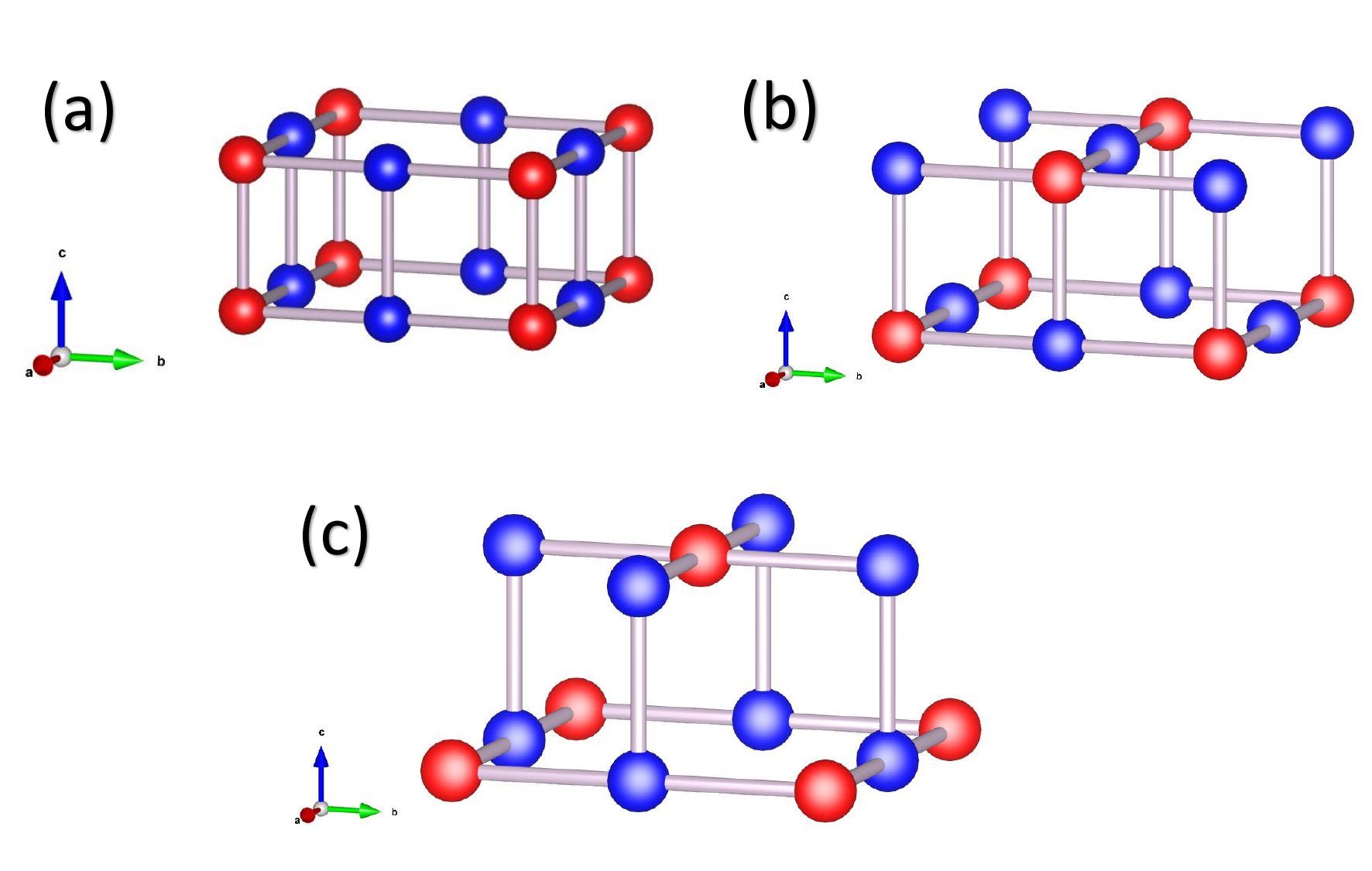}
\caption{(\textbf{a}) $\text{AA}$-stacked bilayer Lieb lattice  (\textbf{b}) $\text{AB}_\text{x}$-stacked bilayer Lieb lattice (\textbf{c}) $\text{AB}_\text{xy}$-stacked bilayer Lieb lattice}
\label{fig:S1}
\end{figure}

\section{The tight-binding model of the generalized Dice lattice}
\par In the basis of $\begin{pmatrix} A, & B, & C\end{pmatrix}^T$, the Bloch Hamiltonian with only nearest-neighbor hopping on a Dice lattice possessing sublattice symmetry is \cite{PhysRevB.103.075418}:
\begin{equation}\label{eq:10}
    H_{\text{bulk}}(\vec{k}) = t\begin{pmatrix} 0 & h & 0 \\ h^* & 0 & h \\ 0 & h^* & 0 \end{pmatrix},
\end{equation}
where $h=1+2e^{-i\frac{3k_y}{2}}cos(\frac{\sqrt{3}k_x}{2})$. Eq.~\ref{eq:09} has triply degenerated touching at $K,K'$ points. If the second nearest-neighbor hopping $t_2$ is considered, the Hamiltonian in Eq.~\ref{eq:10} becomes:
\begin{equation}\label{eq:10}
    H_{\text{bulk}}(\vec{k}) = t\begin{pmatrix} h' & h & 0 \\ h^* & h' & h \\ 0 & h^* & h' \end{pmatrix},
\end{equation}
where $h'=\frac{2t_2}{t}(\cos(\frac{\sqrt{3}}{2}k_x+\frac{3}{2}k_y)+\cos(\frac{\sqrt{3}}{2}k_x-\frac{3}{2}k_y)+\cos(\sqrt{3}k_x))$. The low-energy state around the valleys is described by:
\begin{equation}\label{eq:11}
    H_0(\vec{k}) = \frac{3t}{2}(\pm k_xS_1 - k_yS_2)+\frac{9t_2}{4}(k_x^2+k_y^2),
\end{equation}
where $S$ is the spin-$1$ representation of $SU(2)$.

\begin{figure}[ht]
\centering
\includegraphics[width=8.6cm, height=5cm]{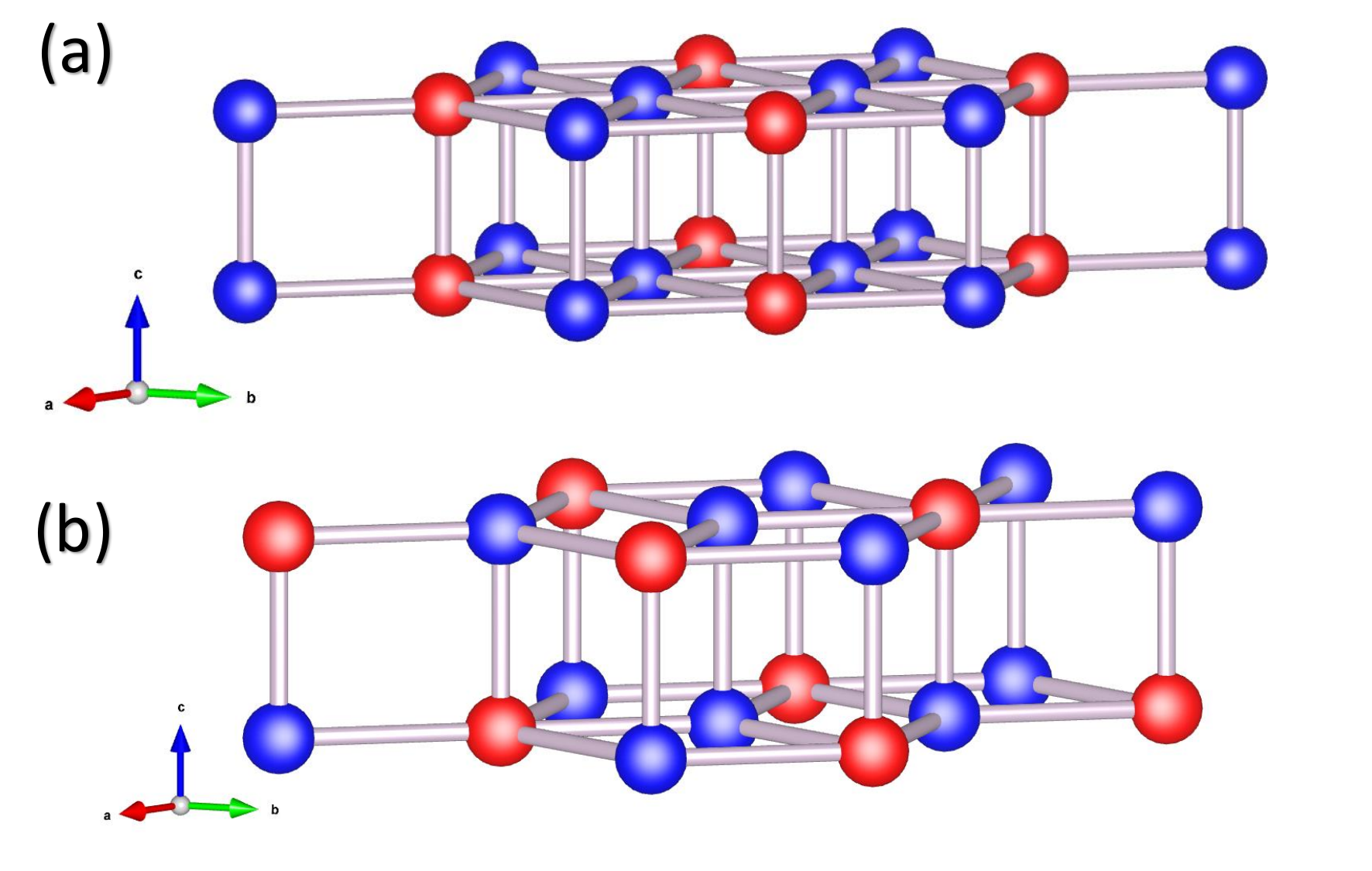}
\caption{(\textbf{a}) $\text{AA}$-stacked bilayer Lieb lattice  (\textbf{b}) $\text{AB}$-stacked bilayer Lieb lattice}
\label{fig:S2}
\end{figure}
\section{The tight-binding model of the twisted bilayers}
We consider AA-stacked twisted bilayer Lieb(Dice) model. For the Lieb structure, 
any commensurate twist can be described by a socalled 'twist' vector (m, n). The twist angle is determined by $\theta_{(m,n)} = 2 \arctan(m/n)$\cite{can_tummuru_day_elfimov_damascelli_franz_2021}. The Moire structure can be understood as by rotating two perfectly aligned square lattices in opposite directions by $\theta_{(m,n)}/2$. The Moire unit cell contains $3\cdot 2 (m^2 + n^2)$ atomic sites. The two Moire lattice vectors are
\begin{eqnarray}
\bm{T}_1 = \sqrt{(m^2+n^2)}\bm{a}_1 = (\sqrt{(m^2 + n^2)}, 0), \nonumber\\
\bm{T}_2 = \sqrt{(m^2+n^2)}\bm{a}_2 = (0, \sqrt{(m^2 + n^2)}),
\end{eqnarray}
where $\bm{a}_1 = (1, 0), \bm{a}_2 = (0, 1)$ are the lattice vectors of the original square lattice. The interlayer spacing is set as 0.8. We choose (m, n) = (1, 5) in this work.

For each constituent layer, we only consider the hopping between nearest neighbour A-B and B-C (the inter-atomic distance is 0.5), and set the hopping parameter t = 1. 
For the interlayer coupling, we truncate the hopping at the cutoff r=1, so that the hopping within this range is $t_{int}$ = 0.2 while the hopping outside the range is 0. Again, we only consider the hopping A-B and B-C to guaranteed the flatness of the bands.

The commensurate twisted Dice structure can be obtained by rotating the two aligned triangular lattice by $\theta_{i}/2$\cite{PhysRevLett.99.256802}, where
\begin{equation}
\cos(\theta_{i}) = \frac{3i^2+3i+1/2}{3i^2 + 3i + 1}.
\end{equation}
The Moire lattice vectors are 
\begin{eqnarray}
\bm{T}_1 &=& i\bm{a}_1 + (i+1)\bm{a}_2,\nonumber\\
\bm{T}_2 &=& -(i+1)\bm{a}_1 + (2i+1)\bm{a}_2.
\end{eqnarray}
The $\bm{a}_1 = 2.46(\sqrt{3}/2, -1/2)$ and $\bm{a}_2 = 2.46(\sqrt{3}/2, 1/2)$ are the two lattice vectors of single triangular layer. The interlayer spacing is set as 2. We choose i=2 in the calculation.

Similarly as in Lieb structure, for each constituent Dice layer, we only consider the nearest neighbour hopping between A and B, B and C while neglect the hopping between A and C. The intralayer hopping between A and B, B and C is set as 1.
For the interlayer coupling, we truncate the hopping at the cutoff r=2.5. The hopping within this range is $t_{int}$ = 0.2 while the hopping outside the range is 0. Again, we only consider the hopping A-B and B-C to guaranteed the flatness of the bands.

\section{The pseudo-Landau levels in twisted bilayer Dice lattice}
\par In the $t_2=0$ limit, the low-energy states around the valleys of a Dice lattice monolayer are described by:
\begin{equation}\label{eq:S1}
    H_0(\vec{k}) = \frac{3t}{2}(\zeta k_xS_1 - k_yS_2),
\end{equation}
where $S$ is the spin-$1$ representation of $SU(2)$, and $\zeta=\pm$ indicates the valley degree of freedom. On the other hand, the interlayer coupling in Eq.~4 in the main text expanding to the order of $O(\frac{r}{L_s})$ is \cite{PhysRevB.99.155415}:
\begin{equation}\label{eq:15}
\begin{split}
    T \cong & \omega_2(T_1(1-i\frac{4\pi y}{3L_s})+T_2(1+i\frac{2\pi}{3L_s}(\sqrt{3}x+y))
    \\ & +T_3(1+i\frac{2\pi}{3L_s}(-\sqrt{3}x+y)) + 3\omega_1,
\end{split}
\end{equation}
where $T_n = \cos((n-1)\frac{2\pi}{3})S_1 - \sin((n-1)\frac{2\pi}{3})S_2 + \frac{\omega_3}{\omega_2}(\cos((n-1)\frac{2\pi}{3})\lambda_4+\sin((n-1)\frac{2\pi}{3})\lambda_5)$. Thus, if we substitute the Hamiltonian in Eq.~\ref{eq:S1} into Eq.~3 in the main text, Eq.~\ref{eq:15} can be treated as the effective vector potential in the twisted bilayer Dice lattice, which is $\vec{A}=-\frac{2\pi\omega_2}{L_s}(\zeta y\hat{x}+x\hat{y})$. The Hamiltonian in Eq.~\ref{eq:S1} can thus be written as:
\begin{equation}\label{eq:16}
\begin{split}
    H = & (\frac{3t}{2}\zeta q_x\tau_0-A_x\tau_3)S_1 - (\frac{3t}{2}q_y\tau_0-A_y\tau_3)S_2
     \\ & -\frac{\omega_3}{\omega_2}(A_x\lambda_4+A_y\lambda_5)\tau_3 + 3\omega_1\tau_2.
\end{split}
\end{equation}
The third term in Eq.~\ref{eq:16} couples the Landau levels within each layer and lifts the degeneracy between different layers, while the fourth term couples the Landau levels in different layers.

\par In Landau gauge, the effective vector potential is $\vec{A}=\frac{4\pi\omega_2\zeta}{L_s}x\hat{y}$ and the effective field strength $B=\frac{4\pi\omega_2\zeta}{L_s}$. Then, by introducing the creation(annihilation) operator for Landau level $a^\dagger(a)\equiv\frac{1}{\sqrt{2\hbar eB}}(\pi_x\pm i\pi_y)$ with the canonical momentum $\vec{\pi}\equiv\vec{k}-e\vec{A}$, the Hamiltonian in Eq.~\ref{eq:16} in the chiral limit $\omega_1=\omega_3=0$ can be rewritten as:
\begin{equation}\label{eq:015}
    H = \hbar\omega_c\begin{pmatrix} 0 & a^\dagger & 0 \\ a & 0 & a^\dagger \\ 0 & a & 0 \end{pmatrix},
\end{equation}
where the cyclotron frequency $\omega_c=\frac{3t}{2}\sqrt{\frac{2eB}{\hbar}}$. To solve for the Landau level, we assume the eigenstate of the $n^{th}$ Landau level $\ket{\psi_n(k)}$ with eigenvalue $E_n$ consists of:
\begin{equation}
    \ket{\psi_n(k)} = \frac{e^{ikx}}{\sqrt{2L_xl_B}}(u_n\begin{pmatrix} 1 \\ 0 \\ 0\end{pmatrix}\ket{n+1} + v_n\begin{pmatrix} 0 \\ 1 \\ 0\end{pmatrix}\ket{n} + w_n\begin{pmatrix} 0 \\ 0 \\ 1\end{pmatrix}\ket{n-1}),
\end{equation}
where $u_n$, $v_n$, and $w_n$ are some complex numbers. Here, the magnetic length $l_B=\sqrt{\frac{3L_s\hbar t}{8\pi\omega_2}}$ and the $L_x$ is the system's size in the $\hat{x}$-direction. The eigenstate of the cyclotron motion $\ket{n}$ satisfies $a\ket{n}=\sqrt{n}\ket{n-1}$ and $a^\dagger\ket{n}=\sqrt{n+1}\ket{n+1}$. Then, according to the eigenvalue equation $H\ket{\psi_n(k)}=E_n\ket{\psi_n(k)}$, $E_n$ satisfies:
\begin{equation}\label{eq:B3}
    \text{det}\begin{pmatrix} E_n & -\hbar\omega_c\sqrt{n+1} & 0 \\ -\hbar\omega_c\sqrt{n+1} & E_n & -\hbar\omega_c\sqrt{n} \\ 0 & -\hbar\omega_c\sqrt{n} & E_n\end{pmatrix} = 0.
\end{equation}
Eq.~\ref{eq:B3} has the solution:
\begin{equation}
    E_n = 0 \quad \text{or} \quad E_n  = \lambda\hbar\omega_c\sqrt{2n+1} \quad \forall n\in\mathbb{N},
\end{equation}
leading to a large degeneracy of the zero modes, i.e., $\frac{1}{3}$ positive integer Landau levels are zero modes. In the following, we absorb the index $k$ and the prefactor $\frac{e^{ikx}}{\sqrt{2L_xl_B}}$ into our braket notation for our convenience. Here, $\lambda=\pm1$. The corresponding eigenmodes are:
\begin{equation}
\begin{split}
    \ket{0n} = & \frac{1}{\sqrt{2n+1}}\begin{pmatrix}\sqrt{n}\ket{n+1} \\ 0\ket{n} \\ -\sqrt{n+1}\ket{n-1} \end{pmatrix}
    \\ \ket{\lambda n} = & \frac{\lambda}{\sqrt{4n+2}}\begin{pmatrix}\sqrt{n+1}\ket{n+1} \\ \lambda\sqrt{2n+1}\ket{n} \\ \sqrt{n}\ket{n-1} \end{pmatrix}.
\end{split}
\end{equation}
Then, to calculate the Hall conductivity of these Landau levels, we need to first calculate the matrix elements for the velocity operators:
\begin{align*}
    \bra{0n'}\hat{S}_1\ket{0n} = & 0
    \\ \bra{0n'}\hat{S}_1\ket{\lambda n} = & \frac{1}{\sqrt{4n'+2}}(\sqrt{n'}\delta_{n'+1,n}-\sqrt{n'+1}\delta_{n'-1,n})
    \\ \bra{\lambda'n'}\hat{S}_1\ket{\lambda n} = & \frac{\lambda\lambda'}{2\sqrt{2n+1}\sqrt{2n'+1}}(\lambda\sqrt{n'+1}\sqrt{2n+1}\delta_{n'+1,n}
    \\ & +\lambda'\sqrt{2n'+1}(\sqrt{n+1}\delta_{n',n+1}+\sqrt{n}\delta_{n',n-1})
    \\ & +\lambda\sqrt{n'}\sqrt{2n+1}\delta_{n'-1,n})
    \\ \bra{0n'}\hat{S}_2\ket{0n} = & 0
    \\ \bra{0n'}\hat{S}_2\ket{\lambda n} = & \frac{-i}{\sqrt{4n'+2}}(\sqrt{n'}\delta_{n'+1,n}+\sqrt{n'+1}\delta_{n'-1,n})
    \\ \bra{\lambda'n'}\hat{S}_2\ket{\lambda n} = & \frac{i\lambda\lambda'}{2\sqrt{2n+1}\sqrt{2n'+1}}(-\lambda\sqrt{2n+1}\sqrt{n'+1}\delta_{n'+1,n}
    \\ & +\lambda'\sqrt{2n'+1}(\sqrt{n+1}\delta_{n',n+1}-\sqrt{n}\delta_{n',n-1})
    \\ & +\lambda\sqrt{2n+1}\sqrt{n'}\delta_{n'-1,n})
\end{align*}
The Hall conductivity is calculated by \cite{PhysRevB.99.155415}:
\begin{equation}
\begin{split}
    \sigma(\mu) = \frac{\eta_0e^2}{h} \sum_{kk',N\neq N',\xi\xi'}(\frac{f(E_{N\xi k}-\mu)-f(E_{N'\xi'k'}-\mu)}{(E_{N\xi k}-E_{N'\xi'k'})^2}
    \times \text{Im}(\bra{N\xi k}\hat{v}_x\ket{N'\xi'k'}\bra{N'\xi'k'}\hat{v}_y\ket{N\xi k}))
\end{split}
\end{equation}
where $\xi=\pm1$ or $0$, $\eta_0=-\frac{2\pi\hbar^2}{L_xL_y}$, and $f(E-\mu)$ is the Fermi-Dirac distribution at chemical potential at $\mu$. The Hall conductivity can be rewritten into a form that is more convenient for calculation:
\begin{equation}\label{eq:21}
    \sigma(\mu) = \sigma(\mu)_{\xi=0,\xi'=\lambda} + \sigma(\mu)_{\xi=\lambda,\xi'=0} + \sigma(\mu)_{\xi=\lambda,\xi'=\lambda'},
\end{equation}
where $\lambda,\lambda'=\pm1$. Here,
\begin{equation}\label{eq:22}
\begin{split}
    \sigma(\mu)_{\xi=0,\xi'=\lambda} = & \frac{\eta_0e^2}{h} \sum_{kk',N\neq N',0\lambda}(\frac{f(-\mu)-f(\lambda\hbar\omega_c\sqrt{2N'+1}-\mu)}{\hbar^2\omega_c^2(2N'+1)} \times \text{Im}(\bra{N0 k}\hat{v}_x\ket{N'\lambda k'}\bra{N'\lambda k'}\hat{v}_y\ket{N0 k}))
    \\ \sigma(\mu)_{\xi=\lambda,\xi'=0} = & \frac{\eta_0e^2}{h} \sum_{kk',N\neq N',\lambda 0}(\frac{f(\lambda\hbar\omega_c\sqrt{2N+1}-\mu)-f(-\mu)}{\hbar^2\omega_c^2(2N+1)} \times \text{Im}(\bra{N\lambda k}\hat{v}_x\ket{N'0k'}\bra{N'0k'}\hat{v}_y\ket{N\lambda k}))
    \\ \sigma(\mu)_{\xi=\lambda,\xi'=\lambda'} = & \frac{\eta_0e^2}{h}\sum_{kk',N\neq N',\lambda \lambda'}(\frac{f(\lambda\hbar\omega_c\sqrt{2N+1}-\mu)-f(\lambda'\hbar\omega_c\sqrt{2N'+1}-\mu)}{\hbar^2\omega_c^2(\lambda\sqrt{2N+1}-\lambda'\sqrt{2N'+1})^2} \times \text{Im}(\bra{N\lambda k}\hat{v}_x\ket{N'\lambda'k'}\bra{N'\lambda'k'}\hat{v}_y\ket{N\lambda k})),
\end{split}
\end{equation}
respectively. Since $\hat{v}_i=v_i\hat{S}_i$, Eq.~\ref{eq:22} becomes:
\begin{equation}\label{eq:23}
\begin{split}
    \sigma(\mu)_{\xi=0,\xi'=\lambda} = & \frac{\eta_0e^2}{h} \sum_{k,N\neq N',\lambda}(\frac{f(-\mu)-f(\lambda\hbar\omega_c\sqrt{2N'+1}-\mu)}{\hbar^2\omega_c^2(2N'+1)} \times (\frac{v_1v_2}{4N+2}(N\delta_{N+1,N'}-(N+1)\delta_{N-1,N'})))
    \\ \sigma(\mu)_{\xi=\lambda,\xi'=0} = & \frac{\eta_0e^2}{h} \sum_{k,N\neq N',\lambda}(\frac{f(-\mu)-f(\lambda\hbar\omega_c\sqrt{2N+1}-\mu)}{\hbar^2\omega_c^2(2N+1)} \times (\frac{v_1v_2}{4N'+2}(N'\delta_{N'+1,N}-(N'+1)\delta_{N'-1,N})))
    \\ \sigma(\mu)_{\xi=\lambda,\xi'=\lambda'} = & \frac{\eta_0e^2}{h}\sum_{k,N\neq N',\lambda}(\frac{f(\lambda\hbar\omega_c\sqrt{2N+1}-\mu)-f(-\lambda\hbar\omega_c\sqrt{2N'+1}-\mu)}{\hbar^2\omega_c^2(\sqrt{2N+1}+\sqrt{2N'+1})^2} \times \frac{v_1v_2}{4(2N+1)(2N'+1)}
    \\ & \times ((-(N'+1)(2N+1)\delta_{N'+1,N}+(2N'+1)((N+1)\delta_{N',N+1}-N\delta_{N',N-1})+N'(2N+1)\delta_{N'-1,N}))
    \\ = & 0.
\end{split}
\end{equation}
Then, by using the expression in Eq.~\ref{eq:23}, we can rewrite Eq.~\ref{eq:21} into:
\begin{equation}\label{eq:24}
\begin{split}
    \sigma(\mu) = & \frac{v_1v_2}{\hbar^2\omega_c^2}\frac{2\eta_0e^2}{h} \sum_{k,N\neq N',\lambda}(f(-\mu)-f(\lambda\hbar\omega_c\sqrt{2N+3}-\mu))\frac{1}{4N+2}\frac{N}{2N+3} 
    \\ & - (f(-\mu)-f(\lambda\hbar\omega_c\sqrt{2N+1}-\mu))\frac{1}{2N+1}\frac{N+2}{4N+6}
    \\ = & \begin{cases} -\frac{e^2}{3h} & \text{, if } 0<\mu<\hbar\omega_c \\ \frac{e^2}{3h} & \text{, if } -\hbar\omega_c<\mu<0 \end{cases}.
\end{split}
\end{equation}
Note that Eq.~\ref{eq:24} is the Hall conductivity contributed only from $n\in\mathbb{N}$ Landau levels. To have the total Hall conductivity, we need to consider the contribution from $n=0$ Landau levels:
\begin{equation}
    \begin{cases}
    \ket{0,n=0} = \begin{pmatrix} \ket{0} & 0 & 0 \end{pmatrix}^T & \text{, with } E_{n=0}=0 \\
    \ket{\lambda,n=0} = \frac{1}{\sqrt{2}}\begin{pmatrix} \ket{1} & \lambda\ket{0} & 0 \end{pmatrix}^T & \text{, with } E_{n=0,\lambda}=\lambda\hbar\omega_c.
    \end{cases}
\end{equation}
The only non-vanishing matrix elements of the velocity operators that involve $n=0$ Landau levels are:
\begin{equation}
\begin{split}
    \bra{0,0}\hat{S}_{1(2)}\ket{\lambda,0} = & (-i)\frac{\lambda}{\sqrt{2}} 
    \\ \bra{\lambda',0}\hat{S}_{1(2)}\ket{\lambda,1} = & (-i)(\frac{1}{2}+\frac{\lambda\lambda'}{\sqrt{12}})
    \\ \bra{\lambda',0}\hat{S}_{1(2)}\ket{0,1} = & (-i)\frac{-\lambda'}{\sqrt{3}}
\end{split}
\end{equation}
The corresponding contribution to the Hall conductivity is calculated by:
\begin{equation}\label{eq:20_new}
\begin{split}
    \sigma(\mu) = & \frac{\eta_0e^2}{h}\frac{v_1v_2}{\hbar^2\omega_c^2} \sum_{k,N\neq N',\lambda\lambda'} (f(-\mu)-f(\lambda\hbar\omega_c-\mu))\delta_{\lambda\lambda'} + (f(\lambda'\hbar\omega_c-\mu)-f(\lambda\hbar\omega_c-\mu))(\frac{1}{2}+\frac{\lambda\lambda'}{\sqrt{3}}+\frac{1}{6})
    \\ & + (f(\lambda'\hbar\omega_c-\mu)-f(-\mu))\frac{2}{3}\delta_{\lambda\lambda'}
    \\ = & \begin{cases} \frac{e^2}{3h} & \text{, if } 0<\mu<\hbar\omega_c \\ -\frac{e^2}{3h} & \text{, if } -\hbar\omega_c<\mu<0 \end{cases}.
\end{split}
\end{equation}
Therefore, combining Eq.~\ref{eq:24} and Eq.~\ref{eq:20_new}, the total Hall conductivity contributed from all Landau levels is:
\begin{equation}
    \sigma_{\text{tot}}(\mu) = \begin{cases} 0 & \text{, if } 0<\mu<\hbar\omega_c \\ 0 & \text{, if } -\hbar\omega_c<\mu<0 \end{cases}.
\end{equation}
Therefore, the Chern number of all the zero-mode Landau levels is $C=0$.


\par If the magnetic field is along the $-\hat{z}$ direction, the Hamiltonian in Eq.~\ref{eq:015} becomes:
\begin{equation}
    H = \hbar\omega_c\begin{pmatrix} 0 & a & 0 \\ a^\dagger & 0 & a \\ 0 & a^\dagger & 0 \end{pmatrix},
\end{equation}
and the corresponding eigenmodes are:
\begin{equation}
\begin{split}
    \ket{0n} = & \frac{1}{\sqrt{2n+1}}\begin{pmatrix} -\sqrt{n+1}\ket{n-1} \\ 0\ket{n} \\ \sqrt{n}\ket{n+1} \end{pmatrix}
    \\ \ket{\lambda n} = & \frac{\lambda}{\sqrt{4n+2}}\begin{pmatrix} \sqrt{n}\ket{n-1} \\ \lambda\sqrt{2n+1}\ket{n} \\ \sqrt{n+1}\ket{n+1} \end{pmatrix}
    \\ \ket{0,n=0} = & \begin{pmatrix} 0 \\ 0 \\ \ket{0} \end{pmatrix},\,\ket{\lambda,n=0} = \frac{1}{\sqrt{2}}\begin{pmatrix} 0 \\ \lambda\ket{0} \\ \ket{1} \end{pmatrix}
\end{split}
\end{equation}
The matrix elements of the velocity operators $\hat{v}_y$ flip a sign compared to the system with a magnetic field in the $\hat{z}$ direction. Thus, the Hall conductivity of the zero-mode Landau levels differs by a minus sign from the one under a magnetic field in the $\hat{z}$ direction. 



\end{document}